\documentclass[cup5b]{caps}
\usepackage{graphicx,psfig}
\usepackage{amssymb}
\usepackage{ociwsymp2}  

\HeadText{A. V. Filippenko}

\begin{document}

\pagenumbering{arabic}

\author[]{ALEXEI V. FILIPPENKO\\Department of Astronomy, University of
California, Berkeley}

\chapter{Evidence from Type Ia Supernovae \\ for an Accelerating Universe and 
\\ Dark Energy}

\begin{abstract}

I review the use of Type Ia supernovae (SNe~Ia) for cosmological distance
determinations.  Low-redshift SNe~Ia ($z \lesssim 0.1$) demonstrate that the
Hubble expansion is linear, that $H_0 = 65 \pm 2$ (statistical) km s$^{-1}$
Mpc$^{-1}$, and that the properties of dust in other galaxies are similar to
those of dust in the Milky Way.  The light curves of high-redshift ($z =
0.3$--1) SNe~Ia are stretched in a manner consistent with the expansion of
space; similarly, their spectra exhibit slower temporal evolution (by a factor
of $1 + z$) than those of nearby SNe~Ia.  The measured luminosity distances of
SNe~Ia as a function of redshift have shown that the expansion of the Universe
is currently accelerating, probably due to the presence of repulsive dark
energy such as Einstein's cosmological constant ($\Lambda$).  Combining our
data with existing measurements of the cosmic microwave background (CMB) 
radiation
and with the results of large-scale structure surveys, we find a best fit for
$\Omega_m$ and $\Omega_\Lambda$ of about 0.3 and 0.7, respectively. Other
studies (e.g., masses of clusters of galaxies) also suggest that $\Omega_m
\approx 0.3$. The sum of the densities, $\sim 1.0$, agrees with the value
predicted by most inflationary models for the early Universe: the Universe is
flat on large scales.  A number of possible systematic effects (dust, supernova
evolution) thus far do not seem to eliminate the need for $\Omega_\Lambda >
0$. Most recently, analyses of SNe~Ia at $z = 1.0-1.7$ provide further support
for current acceleration, and give tentative evidence for an early epoch of
deceleration. Current projects include the search for additional SNe~Ia at $z >
1$ to confirm the early deceleration, and the measurement of a few hundred
SNe~Ia at $z = 0.2-0.8$ to determine the equation of state of the dark energy,
$w = P/(\rho c^2)$.

\end{abstract}

\section{Introduction}

    Supernovae (SNe) come in two main varieties (see Filippenko 1997b for a
review). Those whose optical spectra exhibit hydrogen are classified as Type
II, while hydrogen-deficient SNe are designated Type I. SNe~I are further
subdivided according to the appearance of the early-time spectrum: SNe~Ia are
characterized by strong absorption near 6150~\AA\ (now attributed to Si~II),
SNe~Ib lack this feature but instead show prominent He~I lines, and SNe~Ic have
neither the Si~II nor the He~I lines. SNe~Ia are believed to result from the
thermonuclear disruption of carbon-oxygen white dwarfs, while SNe~II come from
core collapse in massive supergiant stars. The latter mechanism probably
produces most SNe~Ib/Ic as well, but the progenitor stars previously lost their
outer layers of hydrogen or even helium.

   It has long been recognized that SNe~Ia may be very useful distance
indicators for a number of reasons; see Branch \& Tammann (1992), Branch
(1998), and references therein. (1) They are exceedingly luminous, with peak
$M_B$ averaging $-19.2$ mag if $H_0 = 65$ km s$^{-1}$ Mpc$^{-1}$. (2) ``Normal''
SNe~Ia have small dispersion among their peak absolute magnitudes ($\sigma
\lesssim 0.3$ mag). (3) Our understanding of the progenitors and explosion
mechanism of SNe~Ia is on a reasonably firm physical basis.  (4) Little cosmic
evolution is expected in the peak luminosities of SNe~Ia, and it can be
modeled. This makes SNe~Ia superior to galaxies as distance indicators. (5) One
can perform {\it local} tests of various possible complications and
evolutionary effects by comparing nearby SNe~Ia in different environments.

   Research on SNe~Ia in the 1990s has demonstrated their enormous potential as
cosmological distance indicators. Although there are subtle effects that must
indeed be taken into account, it appears that SNe~Ia provide among the most
accurate values of $H_0$, $q_0$ (the deceleration parameter), $\Omega_m$ (the
matter density), and $\Omega_\Lambda$ [the cosmological constant, $\Lambda
c^2/(3H_0^2)$].

   There have been two major teams involved in the systematic investigation of
high-redshift SNe~Ia for cosmological purposes. The ``Supernova Cosmology
Project'' (SCP) is led by Saul Perlmutter of the Lawrence Berkeley Laboratory,
while the ``High-Z Supernova Search Team'' (HZT) is led by Brian Schmidt of the
Mt. Stromlo and Siding Springs Observatories. I have been privileged to work
with both teams (see Filippenko 2001 for a personal account), but my primary
allegiance is now with the HZT.

\section{Homogeneity and Heterogeneity}

  Until the mid-1990s, the traditional way in which SNe~Ia were used for
cosmological distance determinations was to assume that they are perfect
``standard candles'' and to compare their observed peak brightness with those of
SNe~Ia in galaxies whose distances had been independently determined (e.g.,
with Cepheid variables). The rationale was that SNe~Ia exhibit relatively little
scatter in their peak blue luminosity ($\sigma_B \approx 0.4$--0.5 mag; Branch
\& Miller 1993), and even less if ``peculiar'' or highly reddened objects were
eliminated from consideration by using a color cut.  Moreover, the optical
spectra of SNe~Ia are usually rather homogeneous, if care is taken to compare
objects at similar times relative to maximum brightness (Riess et al. 1997, and
references therein).  Over 80\% of all SNe~Ia discovered through the early
1990s were ``normal'' (Branch, Fisher, \& Nugent 1993).

   From a Hubble diagram constructed with unreddened, moderately distant SNe~Ia
($z \lesssim 0.1$) for which peculiar motions should be small and relative
distances (as given by ratios of redshifts) are accurate, Vaughan et al. (1995)
find that
$$\langle M_B({\rm max})\rangle \ = \ (-19.74 \pm 0.06) + 5\, {\rm log}\, (H_0/50)~{\rm mag}.$$
\noindent 
In a series of papers, Sandage et al. (1996) and Saha et al. (1997)
combine similar relations with {\it Hubble Space Telescope (HST)} Cepheid
distances to the host galaxies of seven SNe~Ia to derive $H_0 = 57 \pm 4$ km
s$^{-1}$ Mpc$^{-1}$.

   Over the past two decades it has become clear, however, that SNe~Ia do {\it
not} constitute a perfectly homogeneous subclass (e.g., Filippenko 1997a,b).
In retrospect this should have been obvious: the Hubble diagram for SNe~Ia
exhibits scatter larger than the photometric errors, the dispersion actually
{\it rises} when reddening corrections are applied (under the assumption that
all SNe~Ia have uniform, very blue intrinsic colors at maximum; van den Bergh
\& Pazder 1992; Sandage \& Tammann 1993), and there are some significant
outliers whose anomalous magnitudes cannot possibly be explained by extinction
alone.

    Spectroscopic and photometric peculiarities have been noted with increasing
frequency in well-observed SNe~Ia. A striking case is SN 1991T; its pre-maximum
spectrum did not exhibit Si~II or Ca~II absorption lines, yet two months past
maximum brightness the spectrum was nearly indistinguishable from that of a
classical SN~Ia (Filippenko et al. 1992b; Phillips et al. 1993).  The light
curves of SN 1991T were slightly broader than the SN~Ia template curves, and
the object was probably somewhat more luminous than average at maximum. Another
well-observed, peculiar SNe~Ia is SN 1991bg (Filippenko et al. 1992a;
Leibundgut et al. 1993; Turatto et al. 1996).  At maximum brightness it was
subluminous by 1.6 mag in $V$ and 2.5 mag in $B$, its colors were intrinsically
red, and its spectrum was peculiar (with a deep absorption trough due to
Ti~II).  Moreover, the decline from maximum was very steep, the $I$-band light
curve did not exhibit a secondary maximum like normal SNe~Ia, and the velocity
of the ejecta was unusually low. The photometric heterogeneity among SNe~Ia is
well demonstrated by Suntzeff (1996) with objects having excellent $BVRI$ light
curves.

\section{Cosmological Uses: Low Redshifts}

   Although SNe~Ia can no longer be considered perfect ``standard candles,''
they are still exceptionally useful for cosmological distance
determinations. Excluding those of low luminosity (which are hard to find,
especially at large distances), most of the nearby SNe~Ia that had been
discovered through the early 1990s were {\it nearly} standard (Branch et
al. 1993; but see Li et al. 2001b for more recent evidence of a higher
intrinsic peculiarity rate).  Also, after many tenuous suggestions (e.g.,
Pskovskii 1977, 1984; Branch 1981), Phillips (1993) found convincing evidence
for a correlation between light curve shape and the luminosity at maximum
brightness by quantifying the photometric differences among a set of nine
well-observed SNe~Ia, using a parameter [$\Delta m_{15}(B)$] that measures the
total drop (in $B$ magnitudes) from $B$-band maximum to $t = 15$ days after $B$
maximum. In all cases the host galaxies of his SNe~Ia have accurate relative
distances from surface brightness fluctuations or from the Tully-Fisher
relation.  The intrinsically bright SNe~Ia clearly decline more slowly than dim
ones, but the correlation is stronger in $B$ than in $V$ or $I$.

   Using SNe~Ia discovered during the Cal\'an/Tololo survey ($z \lesssim 0.1$),
Hamuy et al. (1995, 1996b) confirm and refine the Phillips (1993) correlation
between peak luminosity and $\Delta m_{15}(B)$. Apparently the slope is steep
only at low luminosities; thus, objects such as SN 1991bg skew the slope of the
best-fitting single straight line. Hamuy et al. reduce the scatter in the
Hubble diagram of normal, unreddened SNe~Ia to only 0.17 mag in $B$ and 0.14
mag in $V$; see also Tripp (1997). Yet another parameterization is the
``stretch'' method of Perlmutter et al. (1997) and Goldhaber et al. (2001): the
$B$-band light curves of SNe~Ia appear nearly identical when expanded or
contracted temporally by a factor $(1+s)$, where the value of $s$ varies among
objects. In a similar but distinct effort, Riess, Press, \& Kirshner (1995)
show that the luminosity of SNe~Ia correlates with the detailed {\it shape} of
the overall light curve.

\begin{figure*}[t]
\centerline{\psfig{file=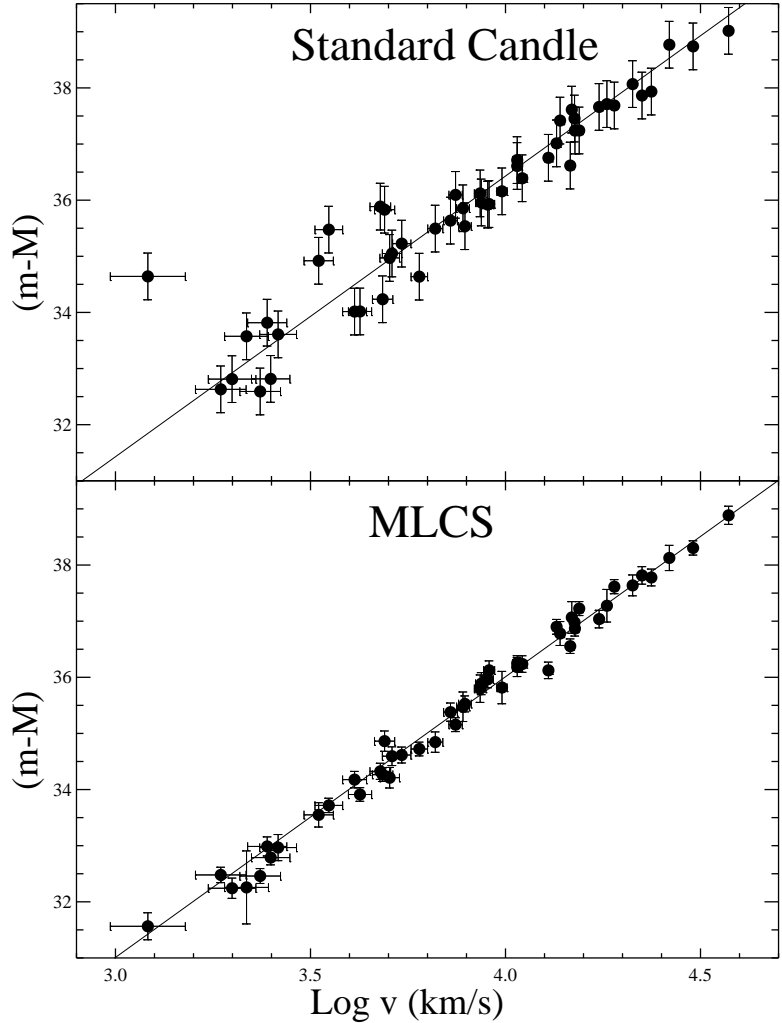,width=11cm,angle=0}}
\hskip 0pt \caption{Hubble diagrams for SNe~Ia (A. G. Riess 2002, private
communication) with velocities (km s$^{-1}$) in the {\it COBE}\ rest frame on
the Cepheid distance scale. The ordinate shows distance modulus, $m - M$ (mag). 
{\it Top:} The objects are assumed to be
{\it standard candles} and there is no correction for extinction; the result
is $\sigma =
0.42$ mag and $H_0 = 58 \pm 8$ km s$^{-1}$ Mpc$^{-1}$. {\it Bottom:} The same
objects, after correction for reddening and intrinsic differences in
luminosity. The result is $\sigma = 0.15$ mag and $H_0 = 65 \pm 2$
(statistical) km s$^{-1}$ Mpc$^{-1}$.
\label{}}
\end{figure*}

   By using light curve shapes measured through several different filters,
Riess, Press, \& Kirshner (1996a) extend their analysis and objectively
eliminate the effects of interstellar extinction: a SN~Ia that has an unusually
red $B-V$ color at maximum brightness is assumed to be {\it intrinsically}
subluminous if its light curves rise and decline quickly, or of normal
luminosity but significantly {\it reddened} if its light curves rise and
decline more slowly. With a set of 20 SNe~Ia consisting of the Cal\'an/Tololo
sample and their own objects, they show that the dispersion decreases from 0.52
mag to 0.12 mag after application of this ``multi-color light curve shape''
(MLCS) method. The results from a recent, expanded set of nearly 50 SNe~Ia
indicate that the dispersion decreases from 0.44 mag to 0.15 mag (Fig. 
1.1). The resulting Hubble constant is $65 \pm 2$ (statistical) km s$^{-1}$
Mpc$^{-1}$, with an additional systematic and zeropoint uncertainty of $\pm 5$
km s$^{-1}$ Mpc$^{-1}$. Riess et al. (1996a) also show that the Hubble flow is
remarkably linear; indeed, SNe~Ia now constitute the best evidence for
linearity. Finally, they argue that the dust affecting SNe~Ia is {\it not} of
circumstellar origin, and show quantitatively that the extinction curve in
external galaxies typically does not differ from that in the Milky Way
(cf. Branch \& Tammann 1992, but see Tripp 1998).

   The advantage of systematically correcting the luminosities of SNe~Ia at
high redshifts rather than trying to isolate ``normal'' ones seems clear in view
of evidence that the luminosity of SNe~Ia may be a function of stellar
population. If the most luminous SNe~Ia occur in young stellar populations
(e.g., Hamuy et al. 1996a, 2000; Branch, Baron, \& Romanishin 1996; Ivanov,
Hamuy, \& Pinto 2000), then we might expect the mean peak luminosity of
high-$z$ SNe~Ia to differ from that of a local sample. Alternatively, the use
of Cepheids (Population I objects) to calibrate local SNe~Ia can lead to a 
zeropoint that is too luminous.  On the other hand, as long as the physics of
SNe~Ia is essentially the same in young stellar populations locally and at high
redshift, we should be able to adopt the luminosity correction methods
(photometric and spectroscopic) found from detailed studies of low-$z$ SNe~Ia.

   Large numbers of nearby SNe~Ia are now being found by my team's Lick Observatory
Supernova Search (LOSS) conducted with the 0.76-m Katzman Automatic Imaging
Telescope (KAIT; Li et al. 2000; Filippenko et al. 2001; see
http://astro.berkeley.edu/$\sim$bait/kait.html). CCD images are taken of $\sim
1000$ galaxies per night and compared with KAIT ``template images'' obtained
earlier; the templates are automatically subtracted from the new images and
analyzed with computer software.  The system reobserves the best candidates the
same night, to eliminate star-like cosmic rays, asteroids, and other sources of
false alarms. The next day, undergraduate students at UC Berkeley examine all
candidates, including weak ones, and they glance at all subtracted images to
locate SNe that might be near bright, poorly subtracted stars or galactic
nuclei. LOSS discovered 20 SNe (of all types) in 1998, 40 in 1999, 38 in 2000,
69 in 2001, and 82 in 2002, making it by far the world's most successful search
for nearby SNe. The most important objects were photometrically monitored
through $BVRI$ (and sometimes $U$) filters (e.g., Li et al. 2001a, 2003; Modjaz
et al. 2001; Leonard et al. 2002a,b), and unfiltered follow-up observations
(e.g., Matheson et al. 2001) were made of most of them during the course of the
SN search. This growing sample of well-observed SNe~Ia should allow us to more
precisely calibrate the MLCS method, as well as to look for correlations
between the observed properties of the SNe and their environment (Hubble type
of host galaxy, metallicity, stellar population, etc.).

\section{Cosmological Uses: High Redshifts}

   These same techniques can be applied to construct a Hubble diagram with
high-redshift SNe~Ia, from which the value of $q_0 = (\Omega_m/2) -
\Omega_\Lambda$ can be determined. With enough objects spanning a range of
redshifts, we can determine $\Omega_m$ and $\Omega_\Lambda$ independently
(e.g., Goobar \& Perlmutter 1995). Contours of peak apparent $R$-band magnitude
for SNe~Ia at two redshifts have different slopes in the
$\Omega_m$--$\Omega_\Lambda$ plane, and the regions of intersection provide the
answers we seek.

\begin{figure*}[t]
\centerline{\psfig{file=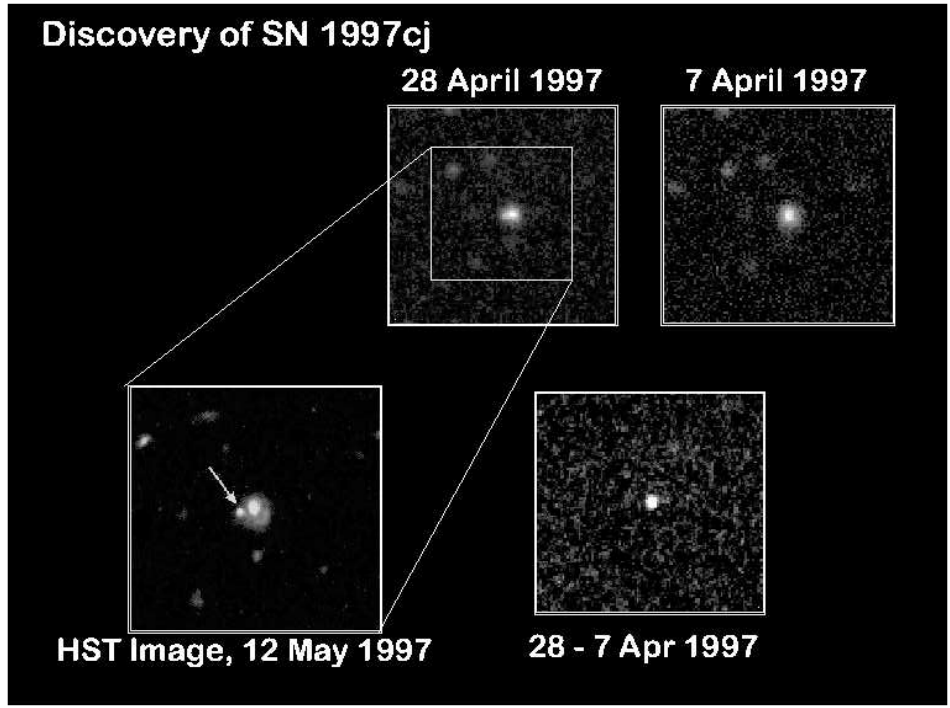,width=13cm,angle=270}}
\hskip 0pt \caption{
Discovery image of SN 1997cj (28 April 1997), along with the
template image and an {\it HST} image obtained subsequently. The net
(subtracted) image is also shown.
\label{}}
\end{figure*}

\subsection{The Search}

   Based on the pioneering work of Norgaard-Nielsen et al. (1989), whose goal
was to find SNe in moderate-redshift clusters of galaxies, the SCP (Perlmutter
et al. 1995a, 1997) and the HZT (Schmidt et al. 1998) devised a strategy that
almost guarantees the discovery of many faint, distant SNe~Ia ``on demand,''
during a predetermined set of nights.  This ``batch'' approach to studying
distant SNe allows follow-up spectroscopy and photometry to be scheduled
in advance, resulting in a systematic study not possible with random
discoveries.  Most of the searched fields are equatorial, permitting follow-up
from both hemispheres.  The SCP was the first group to convincingly demonstrate
the ability to find SNe in batches.

    Our approach is simple in principle. Pairs of first-epoch images are
obtained with wide-field cameras on large telescopes (e.g., the Big Throughput
Camera on the CTIO 4-m Blanco telescope) during the nights around new moon,
followed by second-epoch images 3--4 weeks later.  (Pairs of images permit
removal of cosmic rays, asteroids, and distant Kuiper-belt objects.) These are
compared immediately using well-tested software, and new SN candidates are
identified in the second-epoch images (Fig. 1.2). Spectra are obtained as soon
as possible after discovery to verify that the objects are SNe~Ia and determine
their redshifts.  Each team has already found over 150 SNe in concentrated
batches, as reported in numerous {\it IAU Circulars} (e.g., Perlmutter et
al. 1995b, 11 SNe with $0.16 \lesssim z \lesssim 0.65$; Suntzeff et al. 1996,
17 SNe with $0.09 \lesssim z \lesssim 0.84$). The observed SN~Ia rate at $z
\approx 0.5$ is consistent with the low-$z$ SN~Ia rate together with plausible
star-formation histories (Pain et al. 2002; Tonry et al. 2003), but the error
bars on the high-$z$ rate are still quite large.

   Intensive photometry of the SNe~Ia commences within a few days after
procurement of the second-epoch images; it is continued throughout the ensuing
and subsequent dark runs. In a few cases {\it HST} images are obtained. As
expected, most of the discoveries are {\it on the rise or near maximum
brightness}.  When possible, the SNe are observed in filters that closely
match the redshifted $B$ and $V$ bands; this way, the K-corrections become
only a second-order effect (Kim, Goobar, \& Perlmutter 1996; Nugent, Kim, \&
Perlmutter 2002).  We try to obtain excellent multi-color light curves,
so that reddening and luminosity corrections can be applied (Riess et
al. 1996a; Hamuy et al. 1996a,b).

  Although SNe in the magnitude range 22--22.5 can sometimes be
spectroscopically confirmed with 4-m class telescopes, the signal-to-noise
ratios are low, even after several hours of integration. Certainly Keck or the
VLT are required for the fainter objects (22.5--24.5 mag). With the largest
telescopes, not only can we rapidly confirm a substantial number of candidate
SNe, but we can search for peculiarities in the spectra that might indicate
evolution of SNe~Ia with redshift.  Moreover, high-quality spectra allow us to
measure the age of a SN: we have developed a method for automatically
comparing the spectrum of a SN~Ia with a library of spectra corresponding to
many different epochs in the development of SNe~Ia (Riess et al. 1997).  Our
technique also has great practical utility at the telescope: we can determine
the age of a SN ``on the fly,'' within half an hour after obtaining its
spectrum. This allows us to decide rapidly which SNe are best for subsequent
photometric follow-up, and we immediately alert our collaborators on other
telescopes.

\subsection{Results}

   First, we note that the light curves of high-redshift SNe~Ia are broader
than those of nearby SNe~Ia; the initial indications (Leibundgut et al. 1996;
Goldhaber et al. 1997), based on small numbers of SNe~Ia, are amply confirmed
with the larger samples (Goldhaber et al. 2001). Quantitatively, the amount by
which the light curves are ``stretched'' is consistent with a factor of $1 +
z$, as expected if redshifts are produced by the expansion of space rather than
by ``tired light'' and other non-expansion hypotheses for the redshifts of
objects at cosmological distances. [For non-standard cosmological
interpretations of the SN~Ia data, see Narlikar \& Arp (1997) and Hoyle,
Burbidge, \& Narlikar (2000).]  We also demonstrate this {\it
spectroscopically} at the $2\sigma$ confidence level for a single object: the
spectrum of SN 1996bj ($z = 0.57$) evolved more slowly than those of nearby
SNe~Ia, by a factor consistent with $1 + z$ (Riess et al. 1997). Although one
might be able to argue that something other than universal expansion could be
the cause of the apparent stretching of SN~Ia light curves at high redshifts,
it is much more difficult to attribute apparently slower evolution of spectral
details to an unknown effect.

   The formal value of $\Omega_m$ derived from SNe~Ia has changed with time.
The SCP published the first result (Perlmutter et al. 1995a), based on a single
object, SN 1992bi at $z = 0.458$: $\Omega_m = 0.2 \pm 0.6 \pm 1.1$ (assuming
that $\Omega_\Lambda = 0$). The SCP's analysis of their first seven objects
(Perlmutter et al. 1997) suggested a much larger value of $\Omega_m = 0.88 \pm
0.6$ (if $\Omega_\Lambda = 0$) or $\Omega_m = 0.94 \pm 0.3$ (if $\Omega_{\rm
total} = 1$). Such a high-density universe seemed at odds with other,
independent measurements of $\Omega_m$. However, with the subsequent inclusion
of just one more object, SN 1997ap at $z = 0.83$ (the highest known for a SN~Ia
at the time; Perlmutter et al. 1998), their estimates were revised back down to
$\Omega_m = 0.2 \pm 0.4$ if $\Omega_\Lambda = 0$, and $\Omega_m = 0.6 \pm 0.2$
if $\Omega_{\rm total} = 1$; the apparent brightness of SN 1997ap had been
precisely measured with {\it HST}, so it substantially affected the best fits.

   Meanwhile, the HZT published (Garnavich et al. 1998a) an analysis of four
objects (three of them observed with {\it HST}), including SN 1997ck at $z =
0.97$, at that time a redshift record, although they cannot be absolutely certain
that the object was a SN~Ia because the spectrum is too poor. From these data,
the HZT derived that $\Omega_m = -0.1 \pm 0.5$ (assuming $\Omega_\Lambda = 0$)
and $\Omega_m = 0.35 \pm 0.3$ (assuming $\Omega_{\rm total} = 1$), inconsistent
with the high $\Omega_m$ initially found by Perlmutter et al. (1997) but
consistent with the revised estimate in Perlmutter et al. (1998). An
independent analysis of 10 SNe~Ia using the ``snapshot'' distance method (with
which conclusions are drawn from sparsely observed SNe~Ia) gave quantitatively
similar conclusions (Riess et al. 1998a). However, none of these early data
sets carried the statistical discriminating power to detect cosmic
acceleration.

   The SCP's next results were announced at the 1998 January AAS meeting in
Washington, DC. A press conference was scheduled, with the stated purpose of
presenting and discussing the then-current evidence for a low-$\Omega_m$
universe as published by Perlmutter et al. (1998; SCP) and Garnavich et al.
(1998a; HZT). When showing the SCP's Hubble diagram for SNe~Ia, however,
Perlmutter also pointed out tentative evidence for {\it acceleration}! He
stated that the conclusion was uncertain, and that the data were equally
consistent with no acceleration; the systematic errors had not yet been
adequately assessed. Essentially the same conclusion was given by the SCP in
their talks at a conference on dark matter, near Los Angeles, in February 1998
(Goldhaber \& Perlmutter 1998).

\begin{figure*}[t]
\hbox{
\hskip -0.05truein
\vbox{\hsize 2.5 truein
\psfig{figure=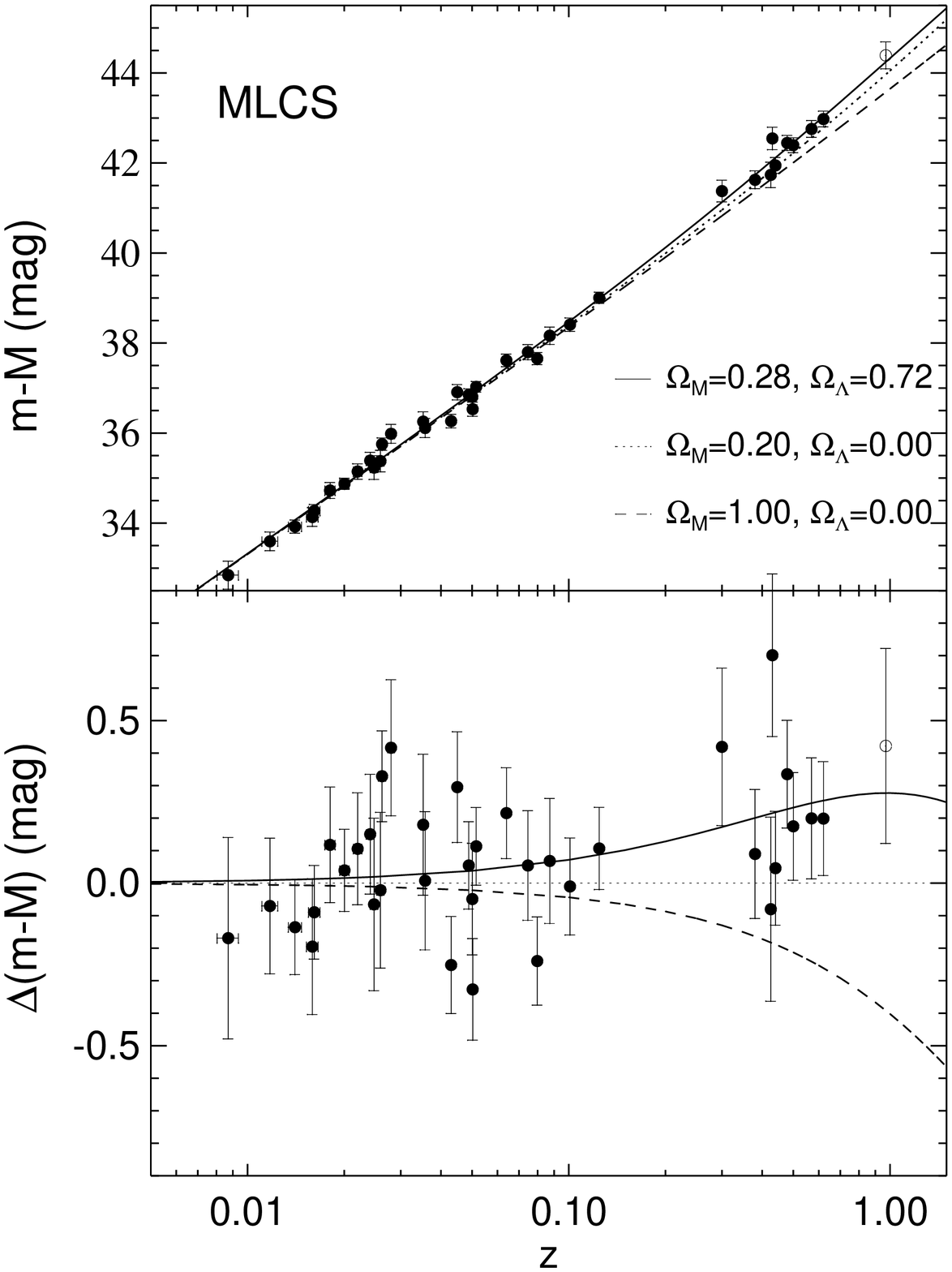,height=3.2truein,angle=0}
}
\hskip +0.0truein
\vbox{\hsize 3.3 truein
\psfig{figure=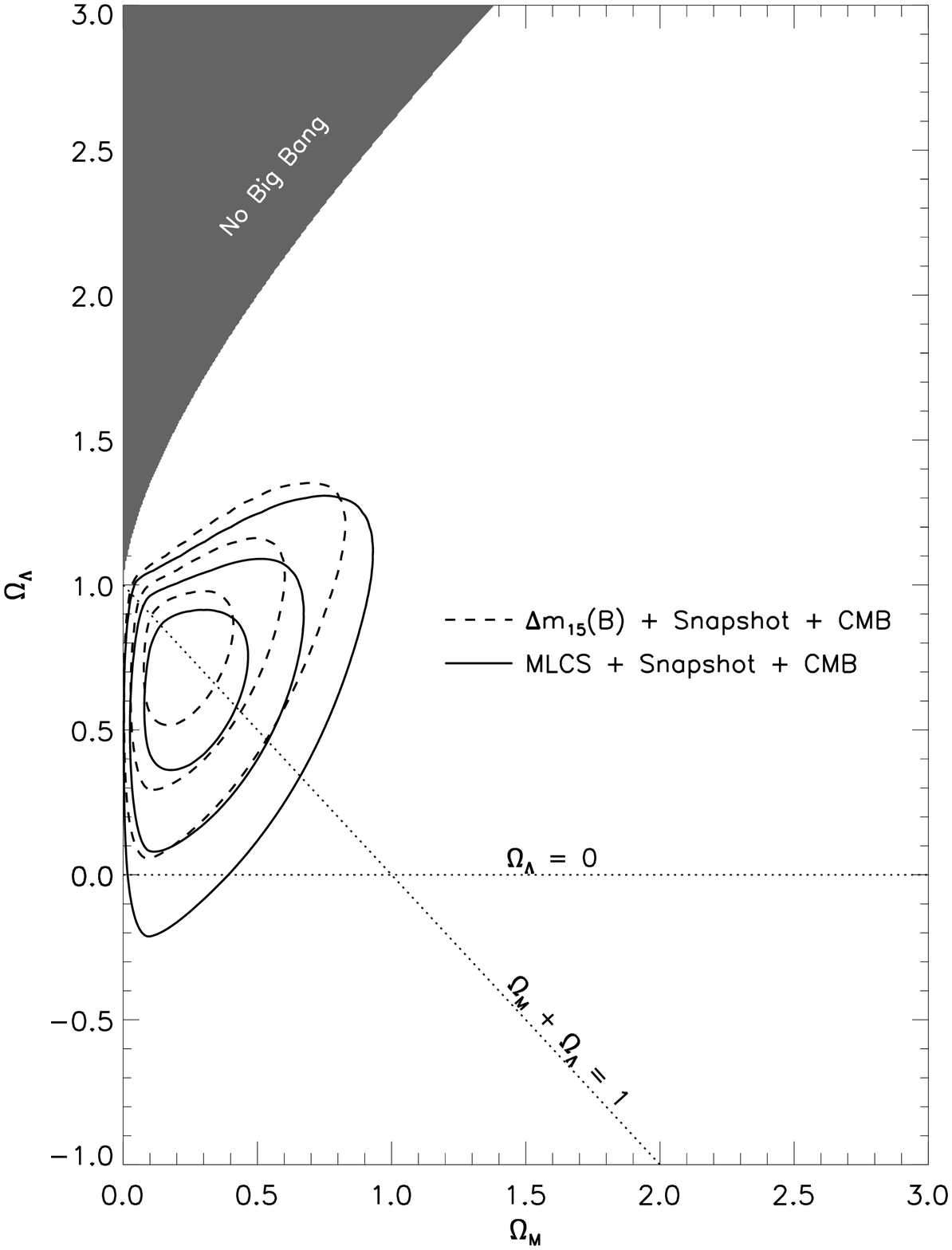,height=3.2truein,angle=0}
}
}
\hskip 0pt \caption{
{\it Left:} The upper panel shows the Hubble diagram for the low-$z$
and high-$z$ HZT SN~Ia sample with MLCS distances; see Riess et
al. (1998b). Overplotted are three world models: ``low'' and ``high''
$\Omega_m$ with $\Omega_\Lambda=0$, and the best fit for a flat universe
($\Omega_m=0.28$, $\Omega_\Lambda=0.72$).  The bottom panel shows the
difference between data and models from the $\Omega_m=0.20$, $\Omega_\Lambda=0$
prediction.  Only the 10 best-observed high-$z$ SNe~Ia are shown.  The average
difference between the data and the $\Omega_m=0.20$, $\Omega_\Lambda=0$
prediction is $\sim 0.25$ mag.
{\it Right:} The HZT's combined constraints from SNe~Ia (left) and the 
position of the first acoustic peak of the CMB angular power spectrum, based 
on data available in mid-1998; see Garnavich et al. (1998b).  The contours 
mark the 68.3\%, 95.4\%, and 99.7\% enclosed probability regions. Solid curves 
correspond to results from the MLCS method; dotted ones are from the 
$\Delta m_{15}(B)$ method; all 16 SNe~Ia in Riess et al. (1998b) were used.
\label{}}
\end{figure*}

   Although it chose not to reveal them at the same 1998 January AAS meeting,
the HZT already had similar, tentative evidence for acceleration in their own
SN~Ia data set. The HZT continued to perform numerous checks of their data
analysis and interpretation, including fairly thorough consideration of various
possible systematic effects. Unable to find any significant problems, even with
the possible systematic effects, the HZT reported detection of a {\it nonzero}
value for $\Omega_\Lambda$ (based on 16 high-$z$ SNe~Ia) at the Los Angeles
dark matter conference in February 1998 (Filippenko \& Riess 1998), and soon
thereafter submitted a formal paper that was published in September 1998 (Riess
et al. 1998b). Their original Hubble diagram for the 10 best-observed high-$z$
SNe~Ia is given in Figure 1.3 ({\it left}). With the MLCS method applied to the 
full set of 16 SNe~Ia, the HZT's formal results were $\Omega_m = 0.24 \pm 
0.10$ if $\Omega_{\rm total} = 1$, or $\Omega_m = -0.35 \pm 0.18$ (unphysical) 
if $\Omega_\Lambda = 0$. If one demanded that $\Omega_m = 0.2$, then the best
value for $\Omega_\Lambda$ was $0.66 \pm 0.21$.  These conclusions did not
change significantly when only the 10 best-observed SNe~Ia were used 
(Fig. 1.3, {\it left}; $\Omega_m = 0.28 \pm 0.10$ if $\Omega_{\rm total} = 1$).

  Another important constraint on the cosmological parameters could be obtained
from measurements of the angular scale of the first acoustic peak of the CMB
(e.g., Zaldarriaga, Spergel, \& Seljak 1997; Eisenstein, Hu, \& Tegmark 1998); 
the SN~Ia and CMB techniques provide nearly complementary information. A 
stunning result was already available by mid-1998 from existing measurements 
(e.g., Hancock et al. 1998; Lineweaver \& Barbosa 1998): the HZT's analysis of 
the SN~Ia data in Riess et al. (1998b) demonstrated that
$\Omega_m + \Omega_\Lambda = 0.94 \pm 0.26$ (Fig. 1.3, {\it right}), when the 
SN and CMB constraints were combined (Garnavich et al. 1998b; see also 
Lineweaver 1998, Efstathiou et al. 1999, and others). 

   Somewhat later (June 1999), the SCP published almost identical results,
implying an accelerating expansion of the Universe, based on an essentially
independent set of 42 high-$z$ SNe~Ia (Perlmutter et al. 1999). Their data,
together with those of the HZT, are shown in Figure 1.4 ({\it left}), and the 
corresponding confidence contours in the $\Omega_\Lambda$ vs.  
$\Omega_m$ plane are given in Figure 1.4 ({\it right}). This incredible 
agreement suggested that neither group had made a
large, simple blunder; if the result was wrong, the reason must be subtle.  Had
there been only one team working in this area, it is likely that far fewer
astronomers and physicists throughout the world would have taken the result
seriously.

   Moreover, already in 1998--1999 there was tentative evidence that the ``dark
energy'' driving the accelerated expansion was indeed consistent with the
cosmological constant, $\Lambda$. If $\Lambda$ dominates, then the equation of
state of the dark energy should have an index $w = -1$, where the pressure
($P$) and density ($\rho$) are related according to $w = P/(\rho c^2)$.
Garnavich et al. (1998b) and Perlmutter et al. (1999) already set an
interesting limit, $w \lesssim -0.60$ at the 95\% confidence level. However,
more high-quality data at $z \approx 0.5$ are needed to narrow the allowed
range, in order to test other proposed candidates for dark energy such as
various forms of ``quintessence'' (e.g., Caldwell, Dav\'e, \& Steinhardt 1998).

   Although the CMB results appeared reasonably persuasive in 1998--1999, one
could argue that fluctuations on different scales had been measured with
different instruments, and that suble systematic effects might lead to
erroneous conclusions. These fears were dispelled only 1--2 years later, when
the more accurate and precise results of the BOOMERANG collaboration were
announced (de Bernardis et al. 2000, 2002). Shortly thereafter the MAXIMA
collaboration distributed their very similar findings (Hanany et al. 2000;
Balbi et al. 2000; Netterfield et al. 2002; see also the TOCO, DASI, and many
other measurements).  Figure 1.4 ({\it right}) illustrates that the CMB 
measurements tightly constrain $\Omega_{\rm total}$ to be close to unity; we 
appear to live in a flat universe, in agreement with most inflationary models 
for the early
Universe! Combined with the SN~Ia results, the evidence for nonzero
$\Omega_\Lambda$ was fairly strong. Making the argument even more compelling
was the fact that various studies of clusters of galaxies (see summary by
Bahcall et al. 1999) showed that $\Omega_m \approx 0.3$, consistent with the
results in Figures 1.3 and 1.4. Thus, a ``concordance cosmology'' had emerged:
$\Omega_m \approx 0.3$, $\Omega_\Lambda \approx 0.7$ --- consistent with what
had been suspected some years earlier by Ostriker \& Steinhardt (1995; see also
Carroll, Press, \& Turner 1992).

\begin{figure*}[t]
\hbox{
\hskip -0.3truein
\vbox{\hsize 3.0 truein
\psfig{figure=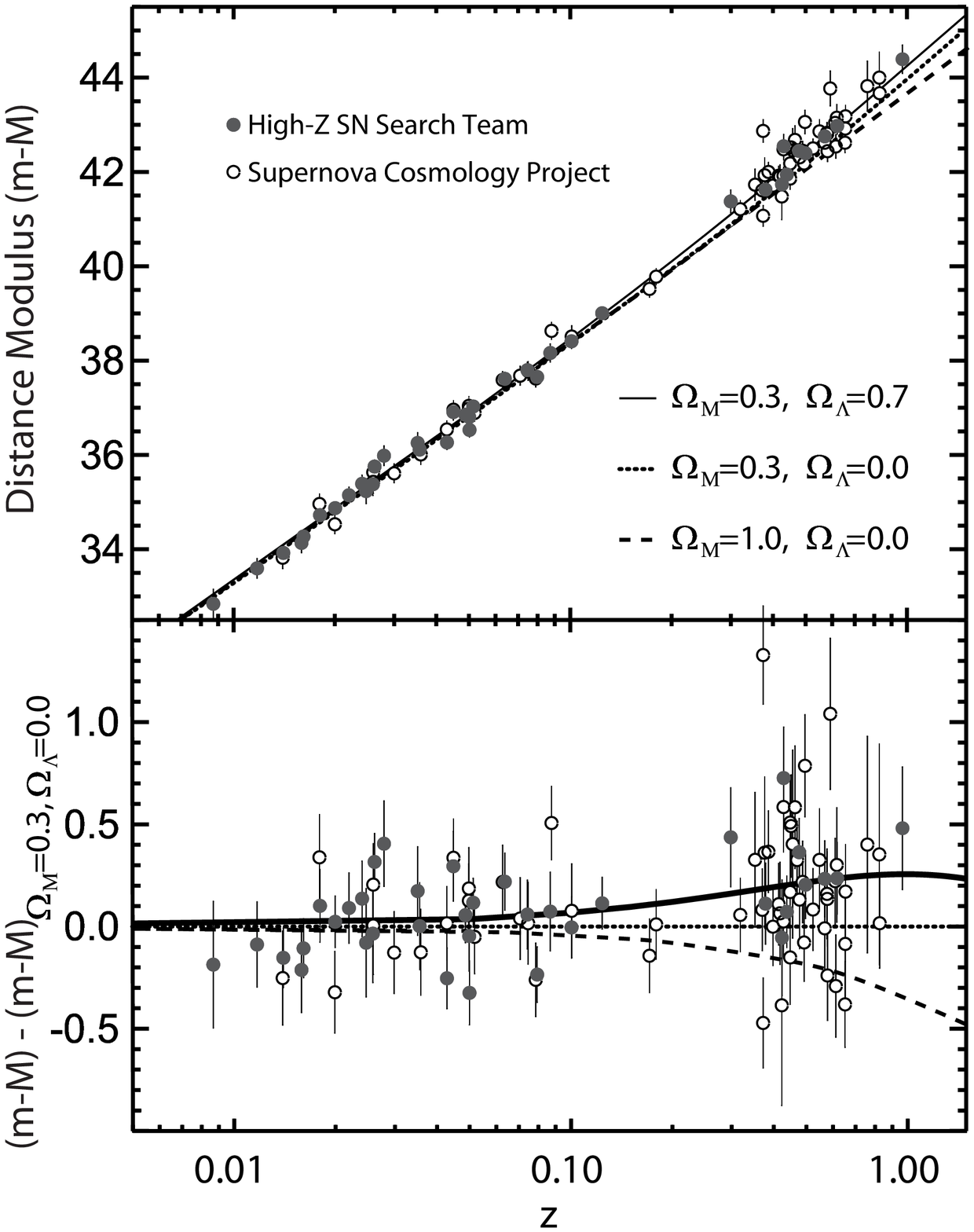,height=3.65truein,angle=0}
}
\hskip -0.2truein
\vbox{\hsize 3.5 truein
\psfig{figure=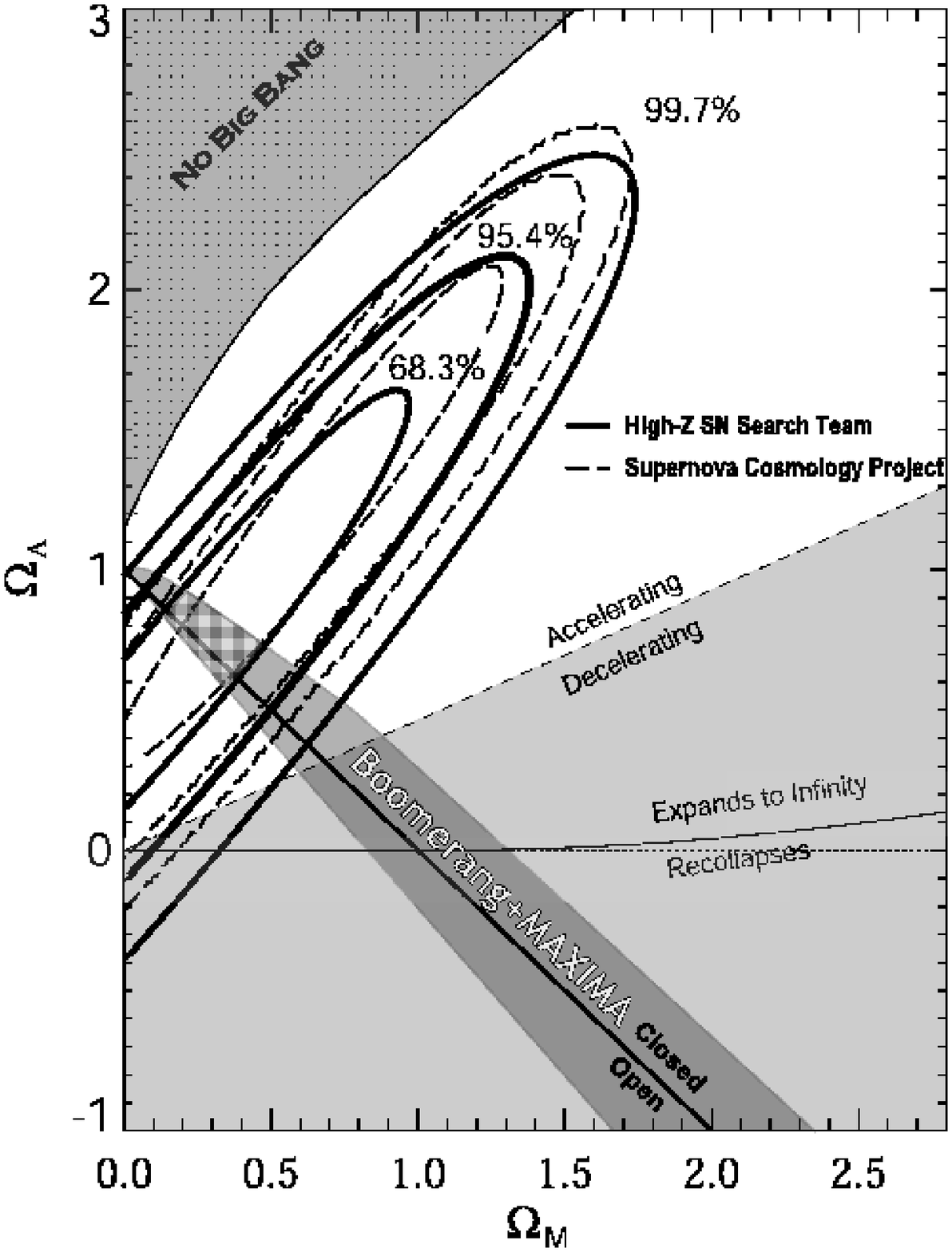,height=3.65truein,angle=0}
}
}
\hskip 0pt \caption{
{\it Left:} As in Fig. 1.3, but this time including both the HZT
(Riess et al. 1998b) and SCP (Perlmutter et al. 1999) samples of low-redshift
and high-redshift SNe~Ia. Overplotted are three world models: $\Omega_m = 0.3$
and $1.0$ with $\Omega_\Lambda=0$, and a flat universe ($\Omega_{\rm total} =
1.0$) with $\Omega_\Lambda = 0.7$.  The bottom panel shows the difference
between data and models from the $\Omega_m=0.3$, $\Omega_\Lambda=0$ prediction.
{\it Right:} The combined constraints from SNe~Ia (left) and the position of
the first acoustic peak of the CMB angular power spectrum, based on BOOMERANG 
and MAXIMA data.  The contours mark the 68.3\%, 95.4\%, and 99.7\% enclosed
probability regions determined from the SNe~Ia. According to the CMB,
$\Omega_{\rm total} \approx 1.0$.
\label{}}
\end{figure*}

  Yet another piece of evidence for a nonzero value of $\Lambda$ was provided
by the Two-Degree Field Galaxy Redshift Survey (2dFGRS; Peacock et al. 2001;
Percival et al. 2001; Efstathiou et al. 2002). Combined with the CMB maps,
their results are inconsistent with a universe dominated by gravitating dark
matter. Again, the implication is that about 70\% of the mass-energy density of
the Universe consists of some sort of dark energy whose gravitational effect is
repulsive. Just as this review was going to press, results from the {\it 
Wilkinson Microwave Anisotropy Prove (WMAP)}\ appeared; together with the 2dFGRS
constraints, they confirm and refine the concordance cosmology ($\Omega_m =
0.27$, $\Omega_\Lambda = 0.73$, $\Omega_{\rm baryon} = 0.044$, $H_0 = 71 \pm 4$
km s$^{-1}$ Mpc$^{-1}$; Spergel et al. 2003).

   The dynamical age of the Universe can be calculated from the cosmological
parameters. In an empty Universe with no cosmological constant, the dynamical
age is simply the ``Hubble time'' (i.e., the inverse of the Hubble constant);
there is no deceleration.  SNe~Ia yield $H_0 = 65 \pm 2$ km s$^{-1}$ Mpc$^{-1}$
(statistical uncertainty only), and a Hubble time of $15.1 \pm 0.5$ Gyr. For a
more complex cosmology, integrating the velocity of the expansion from the
current epoch ($z=0$) to the beginning ($z=\infty$) yields an expression for
the dynamical age. As shown in detail by Riess et al. (1998b), by mid-1998 the
HZT had obtained a value of 14.2$^{+1.0}_{-0.8}$ Gyr using the likely range for
$(\Omega_m, \Omega_\Lambda)$ that they measured.  (The precision was so high
because their experiment was sensitive to roughly the {\it difference} between
$\Omega_m$ and $\Omega_\Lambda$, and the dynamical age also varies in
approximately this way.)  Including the {\it systematic} uncertainty of the
Cepheid distance scale, which may be up to 10\%, a reasonable estimate of the
dynamical age was $14.2 \pm 1.7$ Gyr (Riess et al. 1998b). Again, the SCP's
result was very similar (Perlmutter et al. 1999), since it was based on nearly
the same derived values for the cosmological parameters. This expansion age is
consistent with ages determined from various other techniques such as the
cooling of white dwarfs (Galactic disk $> 9.5$ Gyr; Oswalt et al. 1996),
radioactive dating of stars via the thorium and europium abundances ($15.2 \pm
3.7$ Gyr; Cowan et al.  1997), and studies of globular clusters (10--15 Gyr,
depending on whether {\it Hipparcos} parallaxes of Cepheids are adopted;
Gratton et al. 1997; Chaboyer et al. 1998). By mid-1998, the ages of the oldest
stars no longer seemed to exceed the expansion age of the Universe; the
long-standing ``age crisis'' had evidently been resolved.

\section{Discussion}

   Although the convergence of different methods on the same answer
is reassuring, and suggests that the concordance cosmology is correct, 
it is important to vigorously test each method to make sure it is not
leading us astray. Moreover, only through such detailed studies will
the accuracy and precision of the methods improve, allowing us to eventually
set better constraints on the equation of state parameter, $w$. Here I
discuss the systematic effects that could adversely affect the SN~Ia results.

   High-redshift SNe~Ia are observed to be dimmer than expected in an empty
Universe (i.e., $\Omega_m = 0$) with no cosmological constant. At $z \approx
0.5$, where the SN~Ia observations have their greatest leverage on $\Lambda$,
the difference in apparent magnitude between an $\Omega_m = 0.3$
($\Omega_\Lambda = 0$) universe and a flat universe with $\Omega_\Lambda =0.7$
is only about 0.25 mag. Thus, we need to find out if chemical abundances,
stellar populations, selection bias, gravitational lensing, or grey dust can
have an effect this large. Although both the HZT and SCP had considered many of
these potential systematic effects in their original discovery papers (Riess et
al. 1998b; Perlmutter et al. 1999), and had shown with reasonable confidence
that obvious ones were not greatly affecting their conclusions, if was of
course possible that they were wrong, and that the data were being
misinterpreted.

\subsection{Evolution}

   Perhaps the most obvious possible culprit is {\it evolution} of SNe~Ia over
cosmic time, due to changes in metallicity, progenitor mass, or some other
factor. If the peak luminosity of SNe~Ia were lower at high redshift, then the
case for $\Omega_\Lambda > 0$ would weaken.  Conversely, if the distant
explosions are more powerful, then the case for acceleration strengthens. 
Theorists are not yet sure what the sign of the effect will be, if it is 
present at all; different assumptions lead to different conclusions
(H\"{o}flich et al. 1998; Umeda et al. 1999; Nomoto et al. 2000; 
Yungelson \& Livio 2000).

     Of course, it is extremely difficult, if not effectively impossible, to
obtain an accurate, independent measure of the peak luminosity of high-$z$
SNe~Ia, and hence to directly test for luminosity evolution. However, we can
more easily determine whether {\it other} observable properties of low-$z$ and
high-$z$ SNe~Ia differ. If they are all the same, it is more probable that the
peak luminosity is constant as well --- but if they differ, then the peak
luminosity might also be affected (e.g., H\"{o}flich et al. 1998).  Drell,
Loredo, \& Wasserman (2000), for example, argue that there are reasons to
suspect evolution, because the average properties of existing samples of
high-$z$ and low-$z$ SNe~Ia seem to differ (e.g., the high-$z$ SNe~Ia are more
uniform).

   The local sample of SNe~Ia displays a weak correlation between light curve
shape (or peak luminosity) and host galaxy type, in the sense that the most luminous
SNe~Ia with the broadest light curves only occur in late-type galaxies.  Both
early-type and late-type galaxies provide hosts for dimmer SNe~Ia with narrower
light curves (Hamuy et al.  1996a). The mean luminosity difference for SNe~Ia
in late-type and early-type galaxies is $\sim 0.3$ mag.  In addition, the SN~Ia
rate per unit luminosity is almost twice as high in late-type galaxies as in
early-type galaxies at the present epoch (Cappellaro et al. 1997). These
results may indicate an evolution of SNe~Ia with progenitor age. Possibly
relevant physical parameters are the mass, metallicity, and C/O ratio of the
progenitor (H\"{o}flich et al. 1998).

   We expect that the relation between light curve shape and peak luminosity that
applies to the range of stellar populations and progenitor ages encountered in
the late-type and early-type hosts in our nearby sample should also be
applicable to the range we encounter in our distant sample.  In fact, the range
of age for SN~Ia progenitors in the nearby sample is likely to be {\it larger}
than the change in mean progenitor age over the 4--6 Gyr lookback time to the
high-$z$ sample.  Thus, to first order at least, our local sample should
correct the distances for progenitor or age effects.

   We can place empirical constraints on the effect that a change in the
progenitor age would have on our SN~Ia distances by comparing subsamples of
low-redshift SNe~Ia believed to arise from old and young progenitors.  In the
nearby sample, the mean difference between the distances for the early-type
hosts (8 SNe~Ia) and late-type hosts (19 SNe~Ia), at a given redshift, is 0.04
$\pm$ 0.07 mag from the MLCS method.  This difference is consistent with zero.
Even if the SN~Ia progenitors evolved from one population at low redshift to
the other at high redshift, we still would not explain the surplus in mean
distance of 0.25 mag over the $\Omega_\Lambda=0$ prediction. Moreover, in a
major study of high-redshift SNe~Ia as a function of galaxy morphology, the SCP
found no clear differences (except for the amount of scatter; see \S 1.5.2)
between the cosmological results obtained with SNe~Ia in late-type and
early-type galaxies (Sullivan et al. 2003).

\begin{figure*}[t]
\hbox{
\hskip -0.25truein 
\vbox{\hsize 2.3 truein 
\psfig{figure=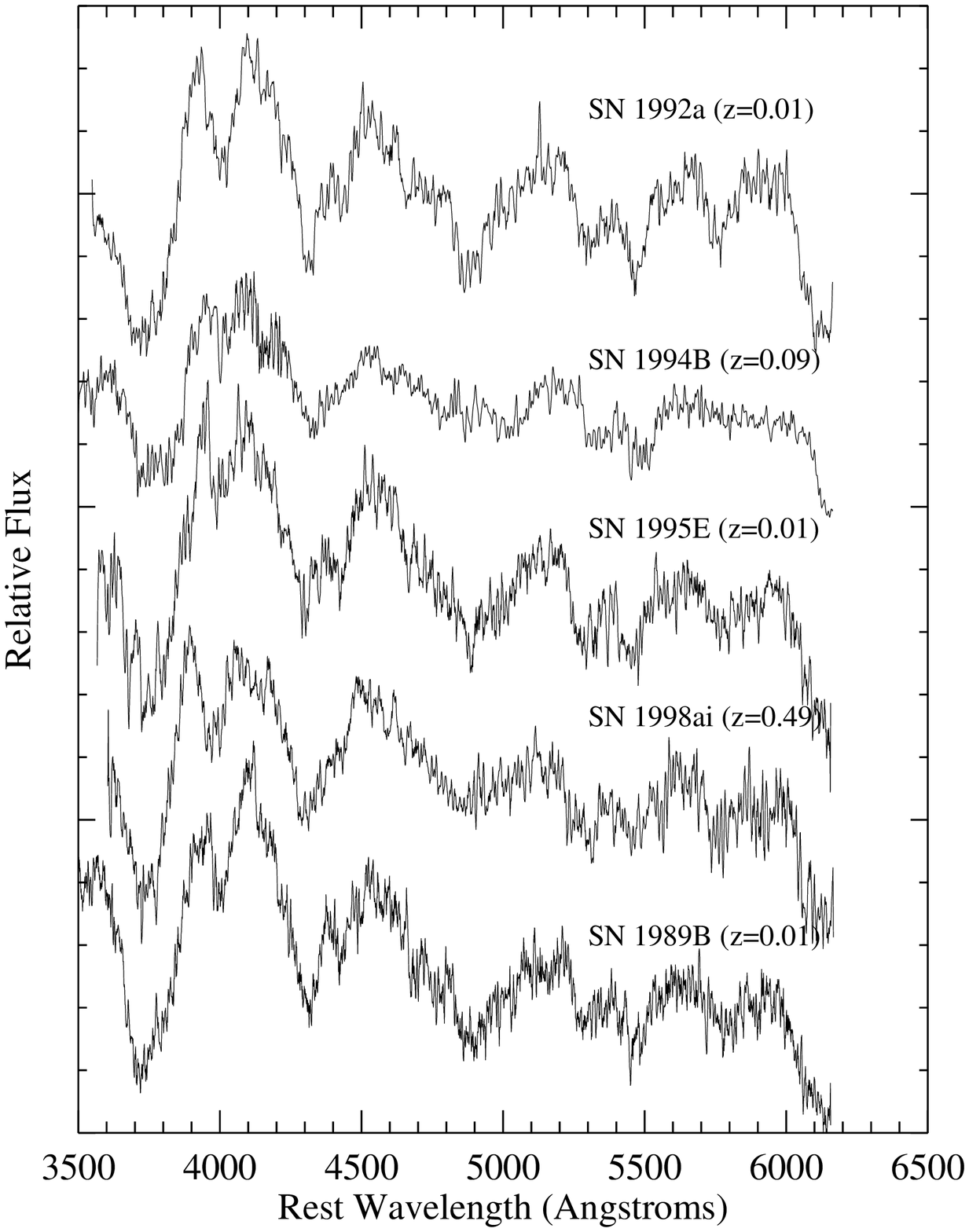,height=3.3truein,angle=0}
}
\hskip 0.0truein 
\vbox{\hsize 2.3 truein 
\psfig{figure=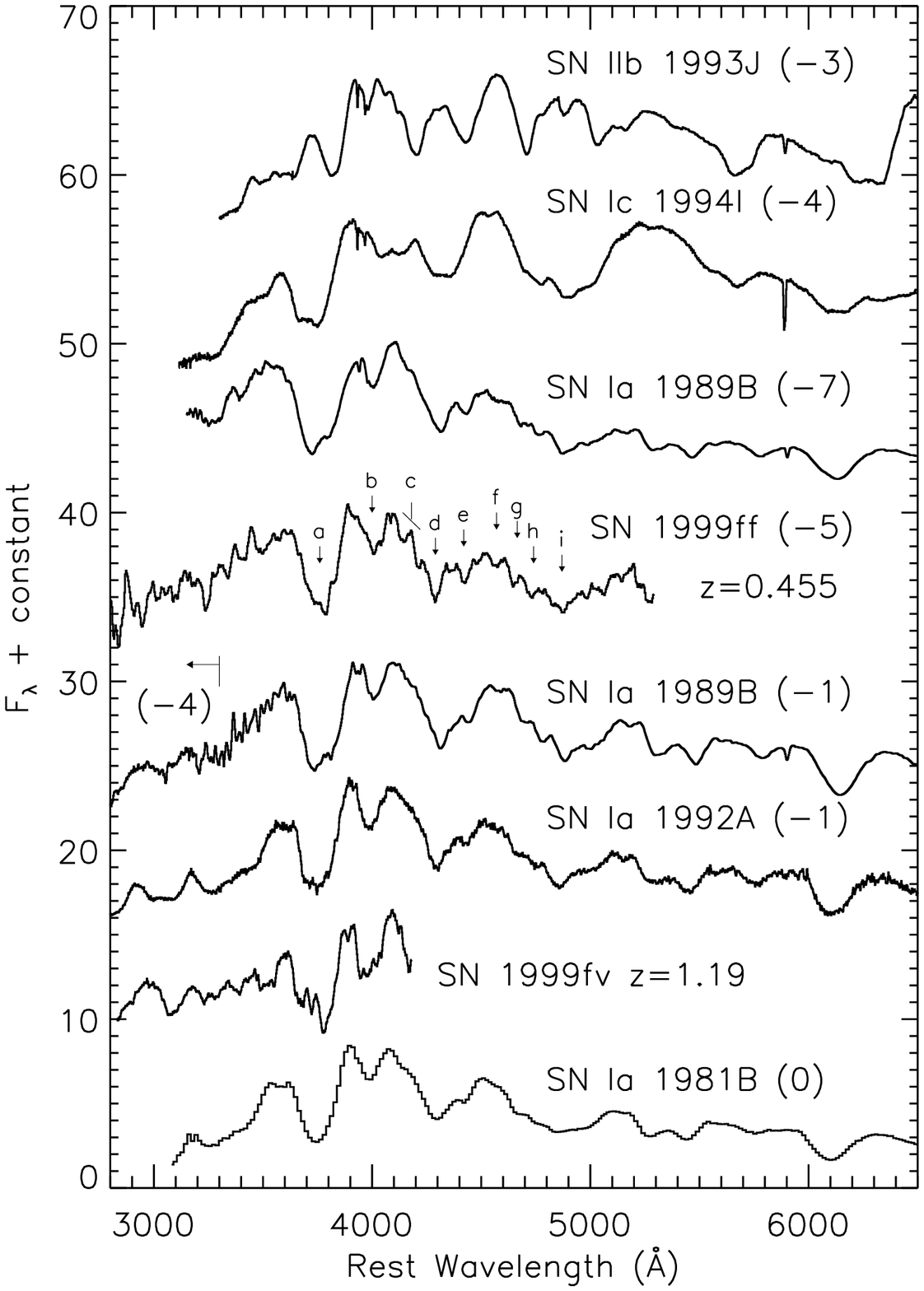,height=3.3truein,angle=0} 
} 
} 
\hskip 0pt \caption{
{\it Left:} 
Spectral comparison (in $f_{\lambda}$) of SN 1998ai ($z = 0.49$; Keck spectrum)
with low-redshift ($z < 0.1$) SNe~Ia at a similar age ($\sim 5$ days before 
maximum brightness), from Riess et al. (1998b).  The spectra of the 
low-redshift SNe~Ia were resampled and convolved with Gaussian noise to match 
the quality of the spectrum of SN 1998ai. Overall, the agreement in the spectra
is excellent, tentatively suggesting that distant SNe~Ia are physically similar 
to nearby SNe~Ia.  SN 1994B ($z = 0.09$) differs the most from the others, and 
was included as a ``decoy.'' 
{\it Right:} 
Heavily smoothed spectra of two high-$z$ SNe (SN 1999ff at $z = 0.455$ and 
SN 1999fv at $z = 1.19$; quite noisy below $\sim$3500~\AA) 
are presented along with several low-$z$ SN Ia spectra (SNe 1989B, 1992A, and 
1981B), a SN Ib spectrum (SN 1993J), and a SN~Ic spectrum (SN 1994I); see 
Filippenko (1997) for a discussion of spectra of various types of SNe. The 
date of the spectra relative to $B$-band maximum is shown in parentheses after 
each object's name. Specific features seen in SN 1999ff and labeled with a letter 
are discussed by Coil et al. (2000). This comparison shows that the two 
high-$z$ SNe are most likely SNe~Ia. 
\label{}}
\end{figure*}

  It is also reassuring that initial comparisons of high-$z$ SN~Ia spectra
appear remarkably similar to those observed at low redshift.  For example, the
spectral characteristics of SN 1998ai ($z = 0.49$) appear to be essentially
indistinguishable from those of normal low-$z$ SNe~Ia; see Figure 1.5 
({\it left}). In fact,
the most obviously discrepant spectrum in this figure is the second one from
the top, that of SN 1994B ($z = 0.09$); it is intentionally included as a
``decoy'' that illustrates the degree to which even the spectra of nearby,
relatively normal SNe~Ia can vary. Nevertheless, it is important to note that a
dispersion in luminosity (perhaps 0.2 mag) exists even among the other, more
normal SNe~Ia shown in Figure 1.5 ({\it left}); thus, our spectra of SN 1998ai 
and other
high-$z$ SNe~Ia are not yet sufficiently good for independent, {\it precise}
determinations of peak luminosity from spectral features (Nugent et al.  1995).
Many of them, however, are sufficient for ruling out other SN types (Fig. 1.5, 
{\it right}),
or for identifying gross peculiarities such as those shown by SNe 1991T and
1991bg; see Coil et al. (2000).

  We can help verify that the SNe at $z \approx 0.5$ being used for cosmology
do not belong to a subluminous population of SNe~Ia by examining restframe
$I$-band light curves. Normal, nearby SNe~Ia show a pronounced second maximum
in the $I$ band about a month after the first maximum and typically about 0.5
mag fainter (e.g., Ford et al. 1993; Suntzeff 1996). Subluminous SNe~Ia, in
contrast, do not show this second maximum, but rather follow a linear decline
or show a muted second maximum (Filippenko et al. 1992a). As discussed by Riess
et al. (2000), tentative evidence for the second maximum is seen from the HZT's
existing $J$-band (restframe $I$-band) data on SN 1999Q ($z = 0.46$); see
Figure 1.6 ({\it left}). Additional tests with spectra and near-infrared light 
curves are currently being conducted.

\begin{figure*}[t]
\hbox{
\hskip -0.35truein
\vbox{\hsize 2.5 truein
\psfig{figure=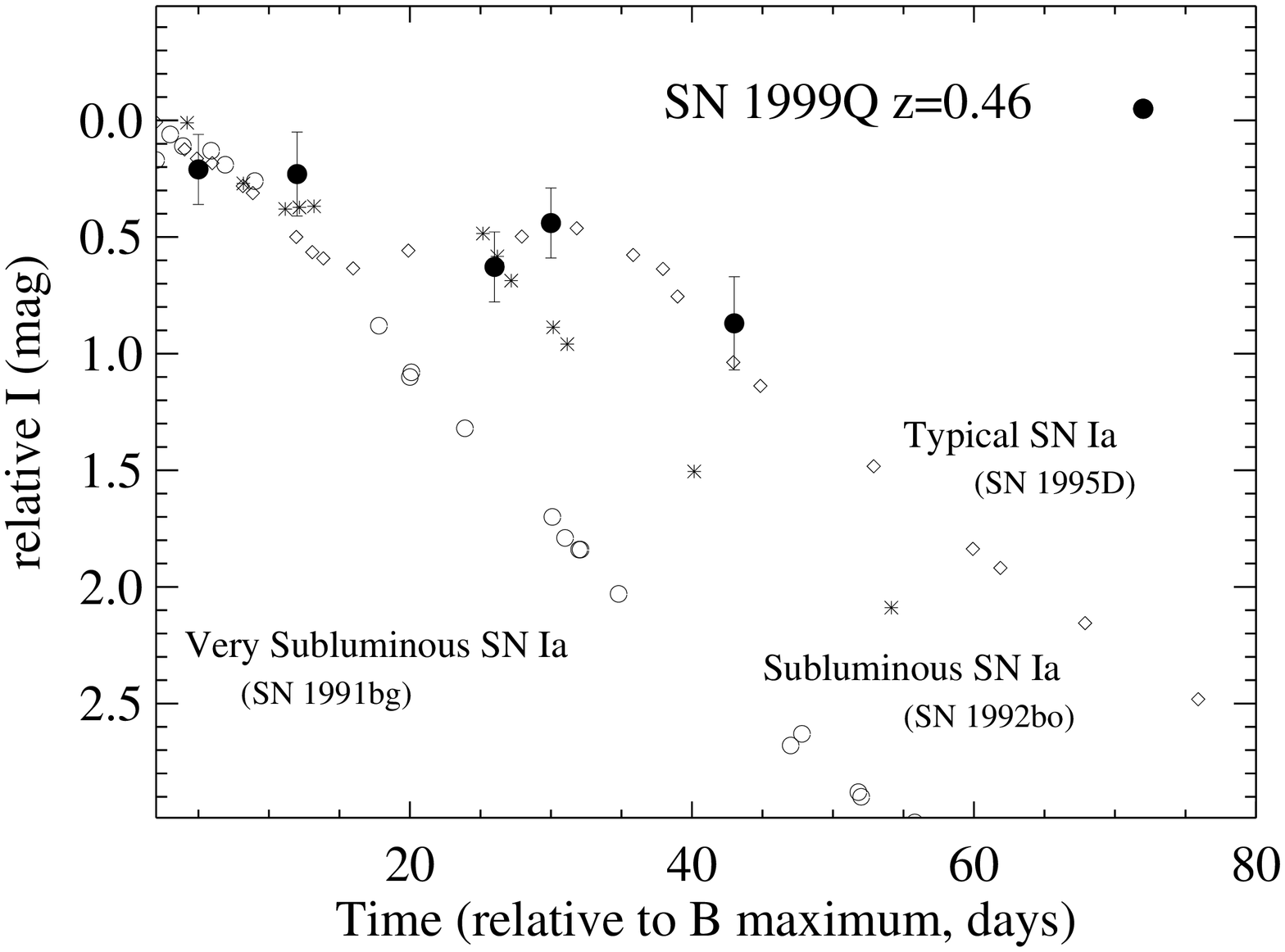,height=2.5truein,angle=90}
}
\hskip -0.0truein
\vbox{\hsize 2.5 truein
\psfig{figure=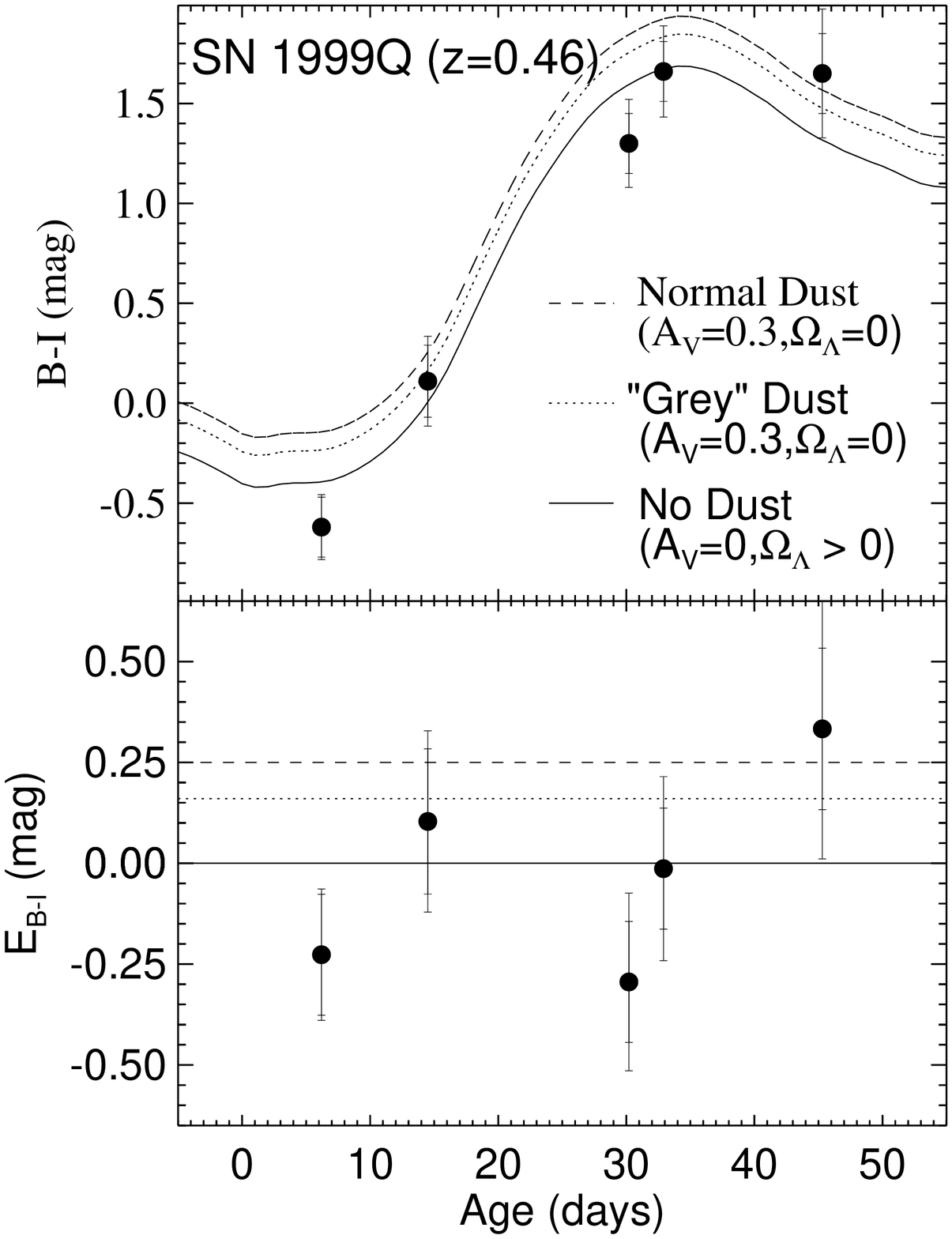,height=2.5truein,angle=0}
}
}
\hskip 0pt \caption{
{\it Left:} Restframe $I$-band (observed $J$-band) light curve of SN
1999Q ($z = 0.46$, 5 solid points; Riess et al. 2000), and the
$I$-band light curves of several nearby SNe~Ia. Subluminous SNe~Ia exhibit
a less prominent second maximum than do normal SNe~Ia.
{\it Right:} Color excess, $E_{B-I}$, for SN 1999Q
and different dust models (Riess et al. 2000). The data are most consistent
with no dust and $\Omega_\Lambda > 0$.
\label{}}
\end{figure*}

  Another way of using light curves to test for possible evolution of SNe~Ia is
to see whether the rise time (from explosion to maximum brightness) is the same
for high-redshift and low-redshift SNe~Ia; a difference might indicate that the
peak luminosities are also different (H\"{o}flich et al. 1998). Riess et al.
(1999c) measured the risetime of nearby SNe~Ia, using data from KAIT, the
Beijing Astronomical Observatory (BAO) SN search, and a few amateur
astronomers. Though the exact value of the risetime is a function of peak
luminosity, for typical low-redshift SNe~Ia it is $20.0 \pm 0.2$ days. Riess et
al. (1999b) pointed out that this differs by $5.8\sigma$ from the {\it
preliminary} risetime of $17.5 \pm 0.4$ days reported in conferences by the SCP
(Goldhaber et al. 1998a,b; Groom 1998). However, more thorough analyses of the
SCP data (Aldering, Knop, \& Nugent 2000; Goldhaber et al. 2001) show that the
high-redshift uncertainty of $\pm 0.4$ days that the SCP originally reported
was much too small because it did not account for systematic effects. The
revised discrepancy with the low-redshift risetime is about $2\sigma$ or
less. Thus, the apparent difference in risetimes might be insignificant. Even
if the difference is real, however, its relevance to the peak luminosity is
unclear; the light curves may differ only in the first few days after the
explosion, and this could be caused by small variations in conditions near the
outer part of the exploding white dwarf that are inconsequential at the peak.

\subsection{Extinction}
 
   Our SN~Ia distances have the important advantage of including corrections
for interstellar extinction occurring in the host galaxy and the Milky
Way. Extinction corrections based on the relation between SN~Ia colors and
luminosity improve distance precision for a sample of nearby SNe~Ia that
includes objects with substantial extinction (Riess et al. 1996a); the scatter
in the Hubble diagram is much reduced.  Moreover, the consistency of the
measured Hubble flow from SNe~Ia with late-type and early-type hosts (see \S
1.5.1) shows that the extinction corrections applied to dusty SNe~Ia at low
redshift do not alter the expansion rate from its value measured from SNe~Ia in
low-dust environments.

   In practice, the high-redshift SNe~Ia generally appear to suffer very little
extinction; their $B-V$ colors at maximum brightness are normal, suggesting
little color excess due to reddening. The most detailed available study is that
of the SCP (Sullivan et al. 2003): they found that the scatter in the Hubble
diagram is minimal for SNe~Ia in early-type host galaxies, but increases for
SNe~Ia in late-type galaxies. Moreover, on average the SNe in late-type
galaxies are slightly fainter (by $0.14 \pm 0.09$ mag) than those in early-type
galaxies. Finally, at peak brightness the colors of SNe~Ia in late-type
galaxies are marginally redder than those in early-type galaxies. Sullivan et
al. (2003) conclude that extinction by dust in the host galaxies of SNe~Ia is
one of the major sources of scatter in the high-redshift Hubble diagram.  By
restricting their sample to SNe~Ia in early-type host galaxies (presumably with
minimal extinction), they obtain a very tight Hubble diagram that suggests a
nonzero value for $\Omega_\Lambda$ at the $5\sigma$ confidence level, under the
assumption that $\Omega_{\rm total} = 1$. In the absence of this assumption,
SNe~Ia in early-type hosts still imply that $\Omega_\Lambda > 0$ at nearly the
98\% confidence level. The results for $\Omega_\Lambda$ with SNe~Ia in
late-type galaxies are quantitatively similar, but statistically less secure
because of the larger scatter.

   Riess, Press, \& Kirshner (1996b) found indications that the Galactic ratios
between selective absorption and color excess are similar for host galaxies in
the nearby ($z \leq 0.1$) Hubble flow.  Yet, what if these ratios changed with
lookback time (e.g., Aguirre 1999a)?  Could an evolution in dust-grain size
descending from ancestral interstellar ``pebbles'' at higher redshifts cause us
to underestimate the extinction?  Large dust grains would not imprint the
reddening signature of typical interstellar extinction upon which our
corrections rely.

   However, viewing our SNe through such gray interstellar grains would also
induce a {\it dispersion} in the derived distances. Using the results of
Hatano, Branch, \& Deaton (1998), Riess et al. (1998b) estimate that the
expected dispersion would be 0.40 mag if the mean gray extinction were 0.25 mag
(the value required to explain the measured MLCS distances without a
cosmological constant).  This is significantly larger than the 0.21 mag
dispersion observed in the high-redshift MLCS distances.  Furthermore, most of
the observed scatter is already consistent with the estimated {\it statistical}
errors, leaving little to be caused by gray extinction.  Nevertheless, if we
assumed that {\it all} of the observed scatter were due to gray extinction, the
mean shift in the SN~Ia distances would be only 0.05 mag.  With the existing
observations, it is difficult to rule out this modest amount of gray
interstellar extinction.

  Gray {\it intergalactic} extinction could dim the SNe without either telltale
reddening or dispersion, if all lines of sight to a given redshift had a
similar column density of absorbing material.  The component of the
intergalactic medium with such uniform coverage corresponds to the gas clouds
producing Lyman-$\alpha$ forest absorption at low redshifts.  These clouds have
individual H~I column densities less than about $10^{15} \, {\rm cm^{-2}}$
(Bahcall et al. 1996).  However, they display low metallicities, typically less
than 10\% of solar. Gray extinction would require larger dust grains which
would need a larger mass in heavy elements than typical interstellar grain size
distributions to achieve a given extinction. It is possible that large dust
grains are blown out of galaxies by radiation pressure, and are therefore not
associated with Lyman-$\alpha$ clouds (Aguirre 1999b).

  But even the dust postulated by Aguirre (1999a,b) and Aguirre \& Haiman
(1999) is not {\it completely} gray, having a size of about 0.1~$\mu$m. We can
test for such nearly gray dust by observing high-redshift SNe~Ia over a wide
wavelength range to measure the color excess it would introduce. If $A_V =
0.25$ mag, then $E(U-I)$ and $E(B-I)$ should be 0.12--0.16 mag (Aguirre
1999a,b). If, on the other hand, the 0.25 mag faintness is due to $\Lambda$,
then no such reddening should be seen.  This effect is measurable using proven
techniques; so far, with just one SN~Ia (SN 1999Q; Fig. 1.6, {\it right}), our 
results favor the no-dust hypothesis to better than 2$\sigma$ (Riess et al. 
2000).  More work along these lines is in progress.

\subsection{The Smoking Gun}

\begin{figure*}[t]
\hbox{
\hskip 0.40truein
\vbox{\hsize 3.8 truein
\vskip -0.9 truein
\psfig{figure=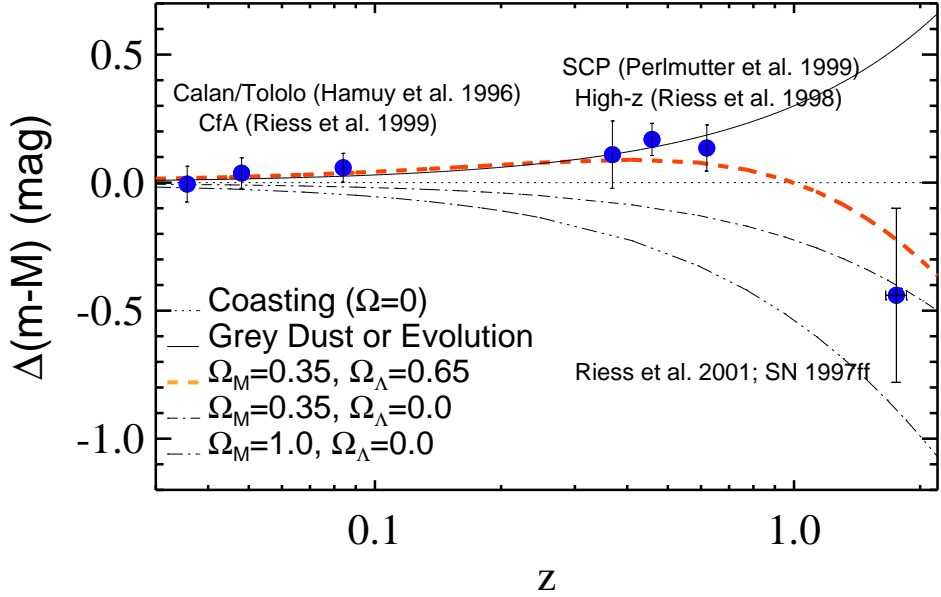,height=4.8truein,angle=0}
\vskip -0.4 truein
}
}
\hskip 0pt \caption{
Hubble diagram for SNe~Ia relative to an empty universe
($\Omega = 0$) compared with cosmological and astrophysical models (Riess et
al. 2001). Low-redshift SNe~Ia are from Hamuy et al. (1996a) and Riess et
al. (1999a). The magnitude of SN 1997ff at $z = 1.7$ has been corrected for
gravitational lensing (Ben\'\i tez et al. 2002). The measurements of SN 1997ff
are inconsistent with astrophysical effects that could mimic previous evidence
for an accelerating universe from SNe~Ia at $z \approx 0.5$.
\label{}}
\end{figure*}

   Suppose, however, that for some reason the dust is {\it very} gray, or our
color measurements are not sufficiently precise to rule out Aguirre's (or
other) dust. Or, perhaps some other astrophysical systematic effect is fooling
us, such as possible evolution of the white dwarf progenitors (e.g.,
H\"{o}flich et al. 1998; Umeda et al. 1999), or gravitational lensing
(Wambsganss, Cen, \& Ostriker 1998). The most decisive test to distinguish
between $\Lambda$ and cumulative systematic effects is to examine the {\it
deviation} of the observed peak magnitude of SNe~Ia from the magnitude expected
in the low-$\Omega_m$, zero-$\Lambda$ model. If $\Lambda$ is positive, the
deviation should actually begin to {\it decrease} at $z \approx 1$; we will be
looking so far back in time that the $\Lambda$ effect becomes small compared
with $\Omega_m$, and the Universe is decelerating at that epoch.  If, on the
other hand, a systematic bias such as gray dust or evolution of the white dwarf
progenitors is the culprit, we expect that the deviation of the apparent
magnitude will continue growing, unless the systematic bias is set up in such
an unlikely way as to mimic the effects of $\Lambda$ (Drell et al. 2000). A
turnover, or decrease of the deviation of apparent magnitude at high redshift,
can be considered the ``smoking gun'' of $\Lambda$.  

   In a wonderful demonstration of good luck and hard work, Riess et al. (2001)
report on {\it HST} observations of a probable SN~Ia at $z \approx 1.7$ (SN
1997ff, the most distant SN ever observed) that suggest the expected turnover
is indeed present, providing a tantalizing glimpse of the epoch of
deceleration. (See also Ben\'\i tez et al. 2002, which corrects the observed
magnitude of SN 1997ff for gravitational lensing.) SN 1997ff was discovered by
Gilliland \& Phillips (1998) in a repeat {\it HST} observation of the Hubble
Deep Field--North, and serendipitously monitored in the infrared with {\it
HST}/NICMOS. The peak apparent SN brightness is consistent with that expected
in the decelerating phase of the concordance cosmological model, $\Omega_m
\approx 0.3$, $\Omega_\Lambda \approx 0.7$ (Fig. 1.7). It is inconsistent with
gray dust or simple luminosity evolution, when combined with the data for
SNe~Ia at $z \approx 0.5$.  On the other hand, it is wise to remain cautious:
the error bars are large, and it is always possible that we are being fooled by
this one object. The HZT and SCP currently have programs to find and measure
more SNe~Ia at such high redshifts. For example, SN candidates at very high
redshifts (e.g., Giavalisco et al. 2002) have been found by ``piggybacking'' on
the Great Observatories Origins Deep Survey (GOODS) being conducted with the
Advanced Camera for Surveys aboard {\it HST}.

  Less ambitious programs, concentrating on SNe~Ia at $z \gtrsim 0.8$, have
already been completed (HZT; Tonry et al. 2003) or are nearing completion
(SCP). Tonry et al. (2003) measured several SNe~Ia at $z \approx 1$, and their
deviation of apparent magnitude from the low-$\Omega_m$, zero-$\Lambda$ model
is roughly the same as that at $z \approx 0.5$, in agreement with expectations
based on the results of Riess et al. (2001). Moreover, the new sample of
high-redshift SNe~Ia presented by Tonry et al., analyzed with methods distinct
from (but similar to) those used previously, confirm the result of Riess et
al. (1998b) and Perlmutter et al. (1999) that the expansion of the Universe is
accelerating. By combining all of the available data sets, Tonry et al. are
able to use 230 SNe~Ia, and they place the following constraints on
cosmological quantities. (1) If the equation of state parameter of the dark
energy is $w = -1$, then $H_0 t_0 = 0.96 \pm 0.04$, and $\Omega_\Lambda - 1.4
\Omega_m = 0.35 \pm 0.14.$ (2) Including the constraint of a flat universe,
they find that $\Omega_m = 0.28 \pm 0.05$, independent of any large-scale
structure measurements. (3) Adopting a prior based on the 2dFGRS constraint on
$\Omega_m$ (Percival et al. 2001) and assuming a flat universe, they derive
that $-1.48 < w < -0.72$ at 95\% confidence.  These constraints are similar in
precision and in value to very recent conclusions reported using {\it WMAP}\
 (Spergel
et al. 2003), also in combination with the 2dFGRS. Complete details on the
SN~Ia results, as well as figures, can be found in Tonry et al. (2003).

\subsection{Measuring the Dark Energy Equation of State}
 
   Every energy component in the Universe can be parameterized by the way its
density varies as the Universe expands (scale factor $a$), with $\rho \propto
a^{-3(1+w)}$, and $w$ reflects the component's equation of state, $w = P/(\rho
c^2)$, where $P$ is the pressure exerted by the component.  So for matter,
$w=0$, while an energy component that does not vary with scale factor has
$w=-1$, as in the cosmological constant $\Lambda$.  Some really strange
energies may have $w < -1$: their density increases with time (Carroll,
Hoffman, \& Trodden 2003)!  Clearly, a good estimate of $w$ becomes the key to
differentiating between models.

The CMB observations imply that the geometry of the universe is close to flat,
so the energy density of the dark component is simply related to the matter
density by $\Omega_x = 1 - \Omega_m$.  This allows the luminosity distance as a
function of redshift to be written as
$$D_L(z)\;
=\; {c(1+z)\over{H_0}}\int_0^{z}{[1+\Omega_x((1+{\rm z})^{3w}-1)]
^{-{1/2}}\over(1+{\rm z})^{{3/2}}}\;  {\rm dz} \; ,$$
showing that the dark energy density and equation of state directly influence
the apparent brightness of standard candles. As demonstrated graphically in
Figure 1.8 ({\it left}), SNe~Ia observed over a wide range of redshifts can 
constrain the dark energy parameters to a cosmologically interesting accuracy.

But there are two major problems with using SNe~Ia to measure $w$.  First,
systematic uncertainties in SN~Ia peak luminosity limit how well $D_L(z)$ can
be measured. While statistical uncertainty can be arbitrarily reduced by
finding thousands of SNe~Ia, intrinsic SN properties such as evolution and
progenitor metallicity, and observational limits like photometric calibrations
and K-corrections, create a systematic floor that cannot be decreased by sheer
force of numbers. We expect that systematics can be controlled to at best 3\%.

Second, SNe at $z > 1.0$ are very hard to discover and study from the
ground. As discussed above, both the HZT and the SCP have found a few SNe~Ia at
$z > 1.0$, but the numbers and quality of these light curves are insufficient
for a $w$ measurement. Large numbers of SNe~Ia at $z > 1.0$ are best left to a
wide-field optical/infrared imager in space, such as the proposed {\it
Supernova/ Acceleration Probe} ({\it SNAP}; Nugent et al. 2000) satellite.

Fortunately, an interesting measurement of $w$ can be made at present.  The
current values of $\Omega_m$ from many methods (most recently {\it WMAP}: 0.27;
Spergel et al. 2003) make an excellent substitute for those expensive SNe at $z
> 1.0$.  Figure 1.8 ({\it left}) shows that a SN~Ia sample with a maximum 
redshift of $z =
0.8$, combined with the current 10\%\ error on $\Omega_m$, will do as well as a
SN~Ia sample at much higher redshifts. Within a few years, the Sloan Digital
Sky Survey and {\it WMAP}\ will solidify the estimate of $\Omega_m$ 
and sharpen $w$ further.

\begin{figure*}[t]
\hbox{
\hskip -0.2truein
\vbox{\hsize 1.9 truein
\psfig{figure=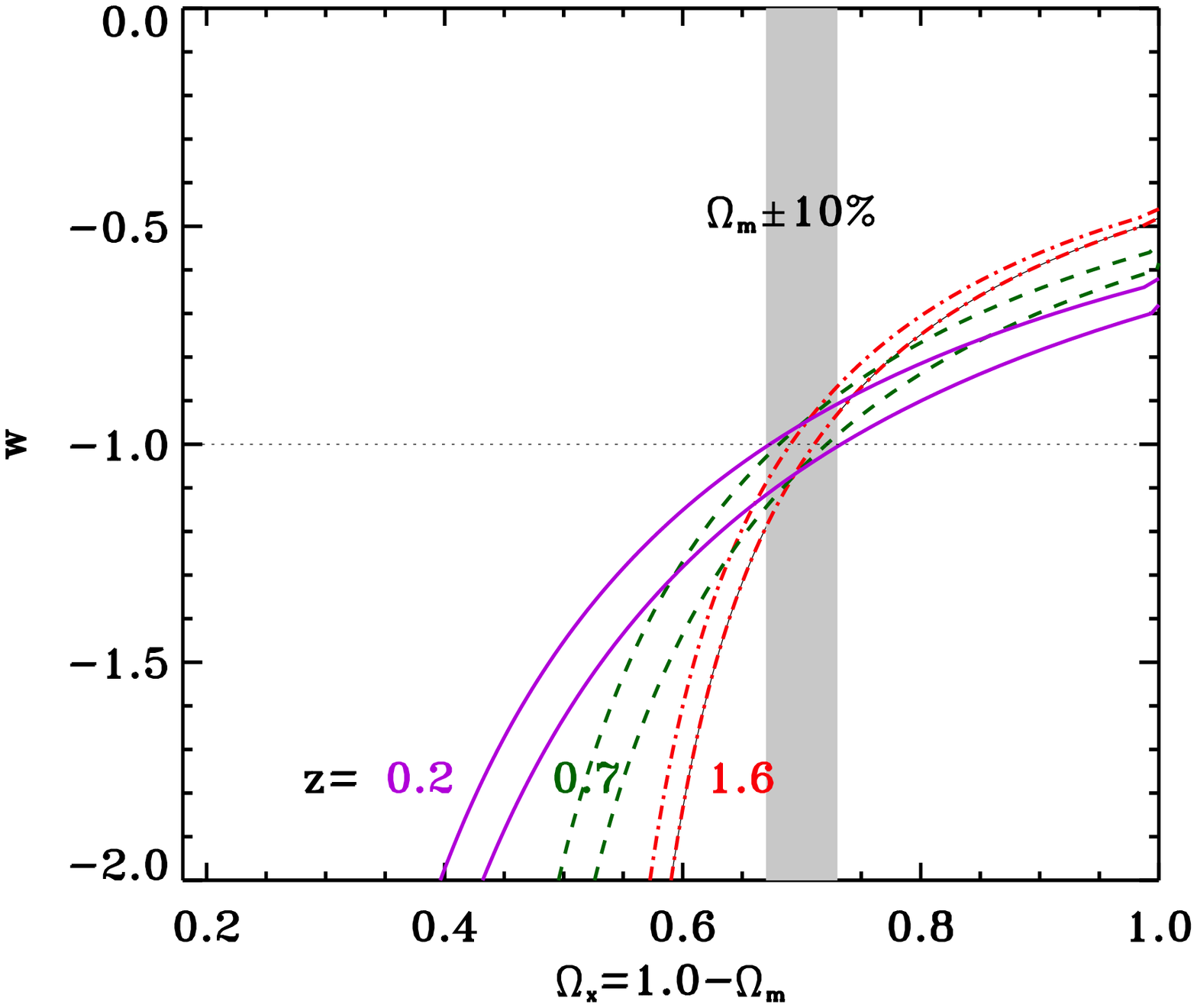,height=1.9truein,angle=0}
}
\hskip +0.0truein
\vbox{\hsize 2.3 truein
\psfig{figure=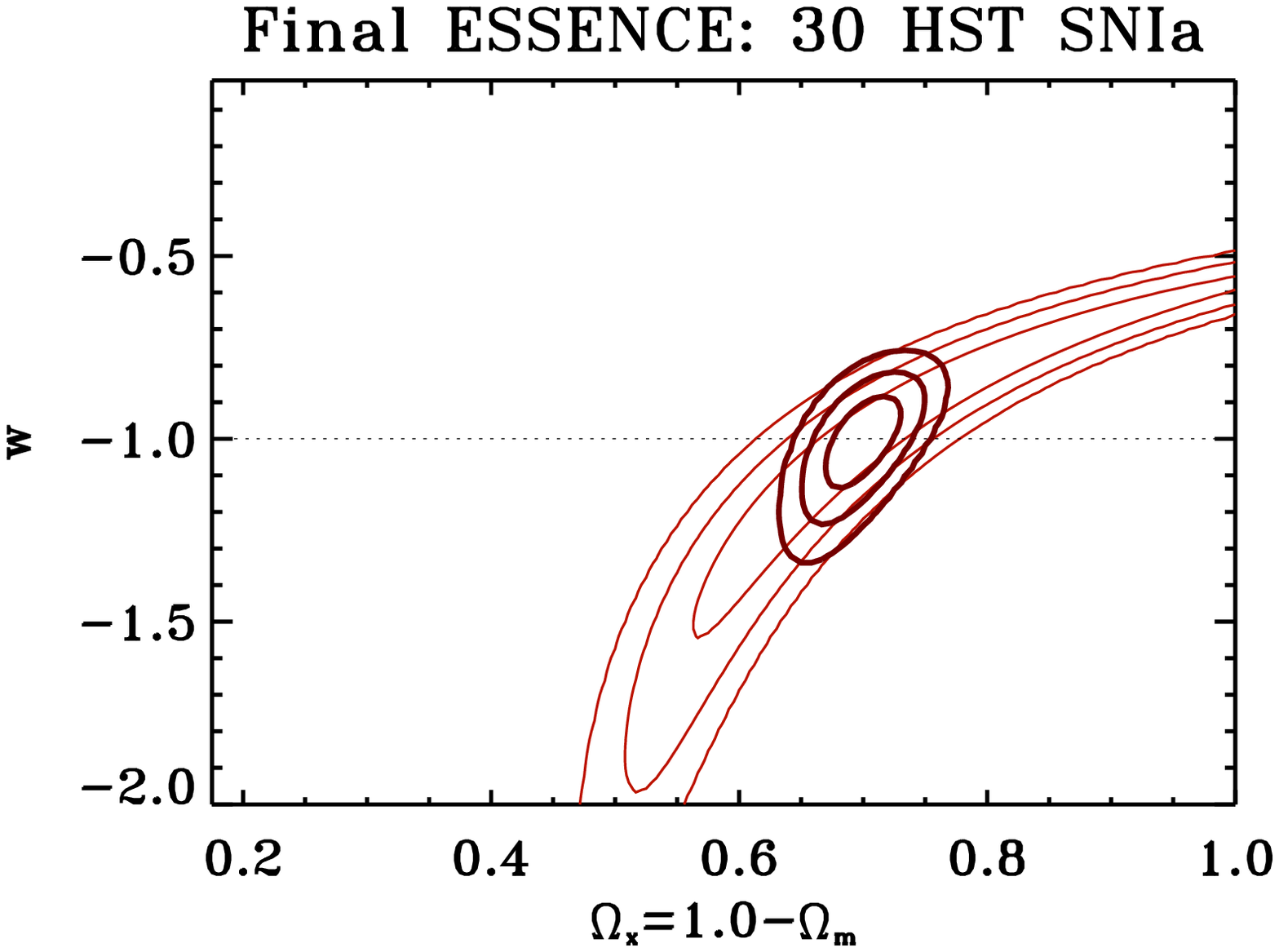,height=1.9truein,angle=0}
}
}
\hskip 0pt \caption{
{\it Left:} Constraints on $\Omega_x$ and $w$ from SN data sets
collected at $z=0.2$ (solid lines), $z=0.7$ (dashed lines), and $z=1.6$
(dash-dot lines). The shaded area indicates how an independent estimate of
$\Omega_m$ with a 10\%\ error can help constrain $w$.
{\it Right:} Expected constraints on $w$ with the desired final
ESSENCE data set of 200 SNe~Ia, 30 of which (in the redshift range $0.6 < z <
0.8$) are to be observed with {\it HST}. The thin lines are for SNe alone while
the thick lines assume an uncertainty in $\Omega_m$ of 7\%. The final ESSENCE
data will constrain the value of $w$ to $\sim$10\%.
\label{}}
\end{figure*}

Both the SCP and the HZT are involved in multi-year programs to discover and
monitor hundreds of SNe~Ia for the purpose of measuring $w$. For example, the
HZT's project, ESSENCE (Equation of State: SupErNovae trace Cosmic Expansion),
is designed to discover 200 SNe~Ia evenly distributed in the $0.2 < z < 0.8$
range.  The CTIO 4-m telescope and mosaic camera are being used to find and
follow the SNe by imaging on every other dark night for several consecutive
months of the year. Keck and other large telescopes are being used to get the
SN spectra and redshifts.  Project ESSENCE will eventually provide an estimate
of $w$ to an accuracy of $\sim$10\% (Fig. 1.8, {\it right}).

   Farther in the future, large numbers of SNe~Ia to be found by the {\it SNAP}
satellite and the Large-area Synoptic Survey Telescope (the ``Dark Matter
Telescope''; Tyson \& Angel 2001) could reveal whether the value of $w$ depends
on redshift, and hence should give additional constraints on the nature of the
dark energy. High-redshift surveys of galaxies such as DEEP2 (Davis et al. 
2001), as well as space-based missions to map the CMB ({\it Planck}), should
provide additional evidence for (or against) $\Lambda$. Observational
cosmology promises to remain exciting for quite some time!

\vspace{0.3cm}
{\bf Acknowledgements}.
  I thank all of my HZT collaborators for their contributions to our team's
research, and members of the SCP for their seminal complementary work on the
accelerating Universe. My group's work at U.C. Berkeley has been supported by
NSF grant AST--9987438, as well as by grants GO--7505, GO/DD--7588, GO--8177,
GO--8641, GO--9118, and GO--9352 from the Space Telescope Science Institute,
which is operated by the Association of Universities for Research in Astronomy,
Inc., under NASA contract NAS~5--26555. Many spectra of high-redshift
SNe were obtained at the W. M. Keck Observatory, which is operated as a 
scientific partnership among the California Institute of Technology, the 
University of California, and NASA; the observatory was made possible by the 
generous financial support of the W. M. Keck Foundation. KAIT has received
donations from Sun Microsystems, Inc., the Hewlett-Packard Company, AutoScope
Corporation, Lick Observatory, the National Science Foundation, the University
of California, and the Sylvia and Jim Katzman Foundation.

\end{document}